\documentclass[
a4paper, 
reprint, 
showkeys, 
nofootinbib,
superscriptaddress, 
twoside,
aps,
prl,
nobibnotes,
twocolumn,
superscriptaddress,
longbibliography]{revtex4-2}

\usepackage[english]{babel}
\addto\captionsenglish{\def\babel@language{en}}
\usepackage[utf8]{inputenc}
\usepackage{amsthm}
\usepackage{mathtools}
\usepackage{physics}
\usepackage{xcolor}
\usepackage{graphicx}
\usepackage[left=23mm,right=13mm,top=35mm,columnsep=15pt]{geometry} 
\usepackage{adjustbox}
\usepackage{placeins}
\usepackage[T1]{fontenc}
\usepackage{lipsum}
\usepackage{csquotes}
\usepackage{hyperref}
\usepackage{soul}
\usepackage{color}
\usepackage{amsmath}
\usepackage{amsfonts}
\usepackage{amssymb}
\usepackage{upgreek}
\usepackage{changes}
\usepackage{marginnote}
\usepackage{comment}
\usepackage{siunitx}

\begin{document}

\title{Ghost Imaging with Free Electron-Photon Pairs}

\author{Sergei Bogdanov\textsuperscript{\dag}}
\affiliation{Vienna Center for Quantum Science and Technology, Atominstitut, TU Wien, Vienna, Austria}
\affiliation{University Service Centre for Transmission Electron Microscopy, TU Wien, Vienna, Austria}

\author{Alexander Preimesberger\textsuperscript{\dag}}
\affiliation{Vienna Center for Quantum Science and Technology, Atominstitut, TU Wien, Vienna, Austria}
\affiliation{University Service Centre for Transmission Electron Microscopy, TU Wien, Vienna, 
Austria}

\author{Harsh Mishra}
\affiliation{Vienna Center for Quantum Science and Technology, Atominstitut, TU Wien, Vienna, Austria}
\affiliation{University Service Centre for Transmission Electron Microscopy, TU Wien, Vienna, Austria}

\author{Dominik Hornof}
\affiliation{Vienna Center for Quantum Science and Technology, Atominstitut, TU Wien, Vienna, Austria}
\affiliation{University Service Centre for Transmission Electron Microscopy, TU Wien, Vienna, Austria}

\author{Thomas Spielauer}
\affiliation{Vienna Center for Quantum Science and Technology, Atominstitut, TU Wien, Vienna, Austria}

\author{Florian Thajer}
\affiliation{Institute of Production Engineering and Photonic Technologies, TU Wien, Vienna, Austria}

\author{Max Maurer}
\affiliation{Vienna Center for Quantum Science and Technology, Atominstitut, TU Wien, Vienna, Austria}
\affiliation{University Service Centre for Transmission Electron Microscopy, TU Wien, Vienna, Austria}

\author{Pia Falb}
\affiliation{Vienna Center for Quantum Science and Technology, Atominstitut, TU Wien, Vienna, Austria}
\affiliation{University Service Centre for Transmission Electron Microscopy, TU Wien, Vienna, Austria}

\author{Leo Stöger}
\affiliation{Vienna Center for Quantum Science and Technology, Atominstitut, TU Wien, Vienna, Austria}
\affiliation{University Service Centre for Transmission Electron Microscopy, TU Wien, Vienna, Austria}

\author{Thomas Schachinger}
\affiliation{University Service Centre for Transmission Electron Microscopy, TU Wien, Vienna, Austria}

\author{Friedrich Bleicher} 
\affiliation{Institute of Production Engineering and Photonic Technologies, TU Wien, Vienna, Austria}

\author{Michael S. Seifner}
\affiliation{Vienna Center for Quantum Science and Technology, Atominstitut, TU Wien, Vienna, Austria}
\affiliation{University Service Centre for Transmission Electron Microscopy, TU Wien, Vienna, Austria}

\author{Isobel C. Bicket}
\affiliation{Vienna Center for Quantum Science and Technology, Atominstitut, TU Wien, Vienna, Austria}
\affiliation{University Service Centre for Transmission Electron Microscopy, TU Wien, Vienna, Austria}

\author{Philipp Haslinger}
\email[]{philipp.haslinger@tuwien.ac.at}
\affiliation{Vienna Center for Quantum Science and Technology, Atominstitut, TU Wien, Vienna, Austria}
\affiliation{University Service Centre for Transmission Electron Microscopy, TU Wien, Vienna, Austria}

\begingroup
\renewcommand{\thefootnote}{\dag}
\footnotetext{These authors contributed equally to this work.}
\endgroup

\date{\today}

\begin{abstract}
Coincidence imaging, also known as ghost imaging, is a technique that exploits correlations between two particles to reconstruct information about a specimen. The particle that relays the spatial information about the object remains completely non-interacting, while the particle used to probe the object is not spatially resolved. 
While ghost imaging has been primarily implemented on photonic platforms, applying it to particles with fundamentally different properties opens up new scientific directions. Mixing massive, charged electrons with massless, neutral photons introduces a new hybrid architecture that unites two fundamental microscopic platforms, each serving as a cornerstone of highly advanced imaging systems.

In this work, we investigate coincidence imaging using electron–cathodoluminescence photon pairs generated within a transmission electron microscope. Utilizing a custom-built free-space cathodoluminescence setup, we demonstrate two-dimensional ghost imaging of complex patterns. We are able to obtain a spatial resolution down to 2 \textmu m, paving the way for adaptation of quantum-enhanced imaging techniques from photonic quantum optics to electron microscopy.
\end{abstract}

\keywords{Coincidence imaging, correlation measurements, cathodoluminescence, transmission electron microscopy, quantum optics}

\maketitle
Over the past decades, imaging methods based on electrons have made significant technological progress, particularly in transmission electron microscopy~(TEM)~\cite{ReimerKohl2008}. A TEM is a highly developed instrument that exploits the wave properties of electrons, combined with advanced electron optics, to resolve and analyse structures at the atomic scale~\cite{ishikawa2023spatial, muller_atomic-scale_2008} with single atom sensitivity~\cite{suenaga_element-selective_2000, varelaSpectroscopicImagingSingle2004}, and the potential to map atomic
orbitals~\cite{lofflerRealspaceMappingElectronic2017,bugnetImagingSpatialDistribution2022}. Temporal resolution in the TEM reaches the sub-nanosecond regime via improvements in direct detector technology~\cite{llopart2022timepix4_electron_detection}, or further down to femto- and even atto-second timescales with the aid of novel pump-probe schemes~\cite{zewail2000femtochemistry, gaida2024attosecond, baum2007attosecond}.
Incorporating coincidence counting techniques into the TEM enables, for example, Hanbury-Brown-Twiss type experiments to probe the correlations of cathodoluminescence (CL) photon emission statistics~\cite{sola-garcia_photon_2021}, to map modes in photonic structures~\cite{feist2022cavity}, or to study excitation pathways~\cite{varkentinaCathodoluminescenceExcitationSpectroscopy2022} and lifetimes~\cite{varkentina2023excitation, yanagimotoTimecorrelatedElectronPhoton2023, meuret_lifetime_2016, meuretTimeresolvedCathodoluminescenceUltrafast2021a}. 
Furthermore, recent advances have demonstrated the ability to sense single CL photon recoils \cite{preimesberger_exploring_2025} and to probe spin dynamics utilizing microwave excitation~\cite{harvey_ultrafast_2021, liu2025correlated, wesels2022beamdeflections}, with angular sensitivities on the order of 100 prad~\cite{jaros2024spin}.
These combined capabilities identify the TEM as an indispensable tool in fields such as materials science, biology, and fundamental research, with expanding possibilities in the study and utilization of electron-light interactions~\cite{abajo_roadmap_2025}.
Within this rapid progress, one avenue that holds great potential, but has so far remained largely unexplored, is to adapt imaging techniques from photonic quantum optics~\cite{pirandola2018advances, moreau2019ghost_imaging, defienne_advances_2024}. Utilizing the quantum properties of photon pairs has allowed researchers to image objects with undetected particles~\cite{Lemos2014ZWM_undetected_photons, Zou1991}, violate the shot noise limit~\cite{slussarenko2017unconditional_entangled}, or use wavefront shaping to address multiple scattering~\cite{yu_recent_2015}, surpassing the performance of classical imaging technologies.
With recent advances in quantum electron microscopy~\cite{Rembold2025_TEM_entanglement, koppell_transmission_2022, kruit_designs_2016, mechel_quantum_2021, shiloh_quantum-coherent_2022, abajo_roadmap_2025, henke_ropers2025entanglement}, such as the first experimental demonstrations of electron-photon entanglement in a TEM~\cite{preimesberger_experimental_2025, henke2025observationquantumentanglementfree}, quantum sensing techniques are now beginning to enter the field of electron microscopy.

Coincidence imaging, also known as ghost imaging, is a technique widely used in photonic quantum optics~\cite{PittmanShihStrekalovSergienko1995, bennink2004ghost_EPR, DAngeloKimKulikShih2004}.
This method uses correlations between two particles to image a specimen located on the path of one particle, while a spatially-resolved detector collects the other particle, which has never interacted with the object~\cite{padgett2017ghost_introduction}. 
A ghost image of the specimen is then formed using coincidence matching between the two particles.
Recently, ghost imaging has been adapted to work with massive ion-ion pairs \cite{khakimov_ghost_2016} and electron-ion pairs \cite{trimeche_ion_2020}.
Working with free electrons, a different form of ghost imaging, known as classical computational ghost imaging, has been demonstrated experimentally in a linear accelerator~\cite{li_electron_2018}, and has been the subject of several proposals for TEM~\cite{kallepalli_challenging_2022, rosi_increasing_2024, rotunno_one-dimensional_2023}.
Nevertheless, quantum ghost imaging, employing entangled electron-photon pairs \cite{preimesberger_experimental_2025} capable of reaching a joint momentum and position resolution beyond the limits of classical physics \cite{mancini_entangling_2002}, has so far not been implemented for the purpose of sensing complex specimens in two dimensions.

The experiment presented herein utilizes the strong spatial correlations present between coherent CL photons and the swift electrons they are emitted from. In contrast to the previous work employing ghost imaging to probe the joint uncertainty relation of position and momentum~\cite{preimesberger_experimental_2025}, the present study focuses on the use of ghost imaging as a tool for complex image reconstruction, demonstrating the first two-dimensional ghost image formed from the spatial correlations between an electron and its corresponding photon. 
It is particularly remarkable that the constituents of these pairs are, in essence, very different, yet both lie at the forefront of microscopic imaging techniques: the charged, massive, relativistic electron possessing a few-picometer de Broglie wavelength, which is able to probe matter on the atomic scale; and the optical photon, which is easy to guide, coherently manipulate, and efficiently detect. 

In the following, we describe the setup and the technique of performing ghost imaging on electron-photon pairs based on a fundamentally different pair-production mechanism and the intrinsic imaging capabilities of a transmission electron microscope, establishing an alternative experimental framework to conventional photon–photon implementations. 
Our work offers an alternative means to produce entangled particle pairs, which could provide several advantages to current principal limitations in enhanced imaging techniques. For example, in photonic quantum ghost imaging, the achievable spatial resolution is necessarily constrained by the nature of entanglement generation in nonlinear crystals, where trade-offs between crystal thickness, phase-matching conditions, pump beam geometry, and entanglement purity impose intrinsic limits on performances~\cite{asban_quantum_2019, moreau_resolution_2018, abouraddy_entangled-photon_2002}. In contrast, the electron–photon platform does not suffer from an analogous fundamental limitation on spatial resolution, due to the interface-localized nature of transition radiation~\cite{frank_transition_1966}, avoiding bulk phase-matching constraints and volume-induced limitations. 
Beyond improved resolution, this platform additionally enables coincidence-based amplitude shaping of electron ~\cite{preimesberger_experimental_2025}, providing access to programmable, post-selected electron states that are not attainable with conventional diffractive or static electron-optical elements~\cite{yu2023AdaptEM}. 

Here, we demonstrate a key feature of the electron-photon platform by imaging a feature-rich, two dimensional pattern in the shape of a cat. In order to benchmark the imaging quality, we also recorded a ghost image of a standard calibration sample, with a spatial resolution down to 2 $\mu$m. 

\begin{figure*}[!t]
	\centering
\includegraphics[width=1\textwidth]{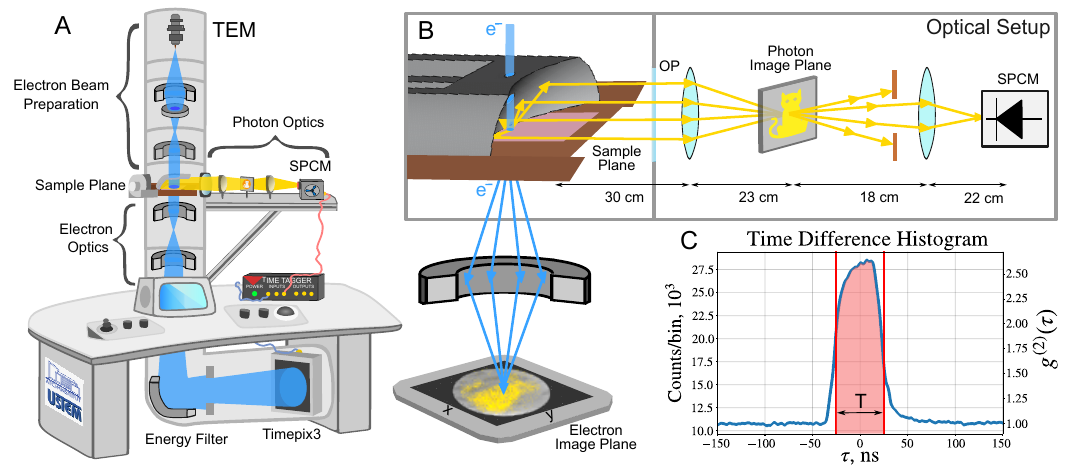} 
	\caption{\textbf{Experimental setup for electron-photon pair coincidence (ghost) imaging}. (\textbf{A}) Schematic of a TEM equipped with an electron energy filter (spectrometer). A 200 keV electron beam interacts with a 50~nm thick silicon membrane, generating correlated electron-CL photon pairs. For ghost imaging, we exploit the strong position correlation of the transition radiation CL photon, which is emitted from the precise point at which the electron enters or exits the membrane. (\textbf{B}) The corresponding CL photons are collected by a parabolic mirror, which is mounted on a single-tilt TEM sample holder with a modified tip. The parabolic mirror and the TEM sample holder are drawn out of scale and enlarged by approximately a factor of 10 for visualization purposes. The photons are then guided through free space via a custom-made optical viewport (OP) to an optical setup. The first lens forms a magnified, spatially-resolved optical image of the electron beam-induced radiation at the TEM's sample plane. To observe the spatial correlations through coincidence imaging, a transmission mask is inserted into this photon image plane. Contributions arising from spherical aberration and surface imperfections of the parabolic mirror are reduced by selecting an optimal collection angle using an aperture. The transmitted photons are subsequently focused onto a single photon counting module (SPCM), which acts as a bucket detector (sensitive only to the counting of photons but lacking spatial resolution), and timestamped. Inside the TEM, operated in low magnification ($510\times$) imaging mode, transmitted electrons are guided through magnetic lenses and energy-filtered to select those that have produced a CL photon. The filtered electrons are then detected with a time-resolved Timepix3-based camera. Each electron detection event is time-stamped. The presented setup ensures that the photon image plane with a transmission mask and the image plane of the electron camera are conjugate planes with respect to the TEM's sample plane. (\textbf{C}) A time tagger ensures the necessary synchronization, enabling the temporal correlation of each individual photon with the electron which emitted it. This correlation is clearly visible in a time difference histogram, which is used to identify electron-photon pairs based on their temporal relationship.
        }
	\label{fig:setup}
\end{figure*}

\section{Experimental setup}

The experiment is conducted using a transmission electron microscope (TEM: Tecnai G2 F20) operated at $E_\mathrm{kin}= 200\; \mathrm{keV}\pm0.45 \;\mathrm{eV}$ (HWHM) with a current of $\sim 23$~pA and adapted for free-space extraction of CL photons, schematically depicted in Fig.~\ref{fig:setup}(A). Electron–photon pairs are generated via the interaction of electrons  with a 50~nm thick monocrystalline silicon membrane (Silson Ltd.). Upon interaction with the sample, electrons may emit CL photons, which predominantly originate from transition radiation~\cite{preimesberger_exploring_2025}.

For transition radiation, the electron and photon are tightly correlated in time and energy. Energy filtering and temporal coincidence matching thus allow us to remove background electrons which have not generated a CL photon, as well as signal originating from incoherent CL and elastically scattered electrons \cite{scheucher2022discrimination, preimesberger_exploring_2025, varkentina2023excitation}. This selection is key to isolating true coherent electron-photon pairs, which possess strong correlations in position space, for the ghost imaging process.

A custom-made parabolic mirror with a focal length of approximately 750 \textmu m, mounted directly on a single tilt TEM sample holder with a modified tip (see Supplemental Material for details), collects these photons. The mirror is precisely aligned to capture photons emitted from the electron beam interaction point and guide them through a retrofitted optical viewport (Thorlabs VC23W4) out of the TEM column. From there, the photons are manipulated by an external optical system, as shown in Fig.~\ref{fig:setup}(B).

The optical setup outside the TEM begins with an achromatic lens (Thorlabs ACA254-150-A) with a focal length of 150~mm, positioned just after the optical viewport. This lens forms a magnified, spatially-resolved optical image of the electron beam-induced radiation at the TEM's sample plane. As a result, photons emitted from a specific location on the silicon membrane are projected onto corresponding locations in the photonic image. 
A cat-shaped mask placed in this image plane selectively transmits or blocks photons based on position, effectively removing absorbed or reflected photons from the electron-photon correlation statistics.
Downstream, an aperture in the image plane of the parabolic mirror selects the optimal collection angle, thus reducing the contributions from spherical aberrations and surface imperfections of the mirror. The parabolic mirror used in this work is an early prototype, and its present surface quality is among the factors that influence the achievable spatial resolution (see Supplemental Material for details). Ongoing progress in the mirror design and fabrication techniques is expected to yield mirrors with improved performance, thereby enhancing the overall resolution of the imaging system. The aberrations could also be mitigated using readily available adaptive optics devices, such as spatial light modulators for arbitrary wavefront shaping~\cite{bowman2010slm} or deformable mirrors~\cite{peck2025performancesimulationskolaachieving}.
Afterwards, a lens with a focal length of 100~mm (Thorlabs ACA254-100-A) collects the transmitted photons and either guides them onto the single-photon counting module (PicoQuant PMA Hybrid 40) or, alternatively, images them onto a low-noise CMOS camera for alignment and diagnostic purposes. In contrast to our earlier work~\cite{preimesberger_experimental_2025}, in the present scheme a bandpass filter was not used, since newly implemented achromatic doublets are already well corrected in the visible spectrum and their residual aberrations do not significantly impact the achievable resolution. All photon detection events are time-stamped using a time-tagging device (Swabian Instruments Time Tagger Ultra).
 
Inside the TEM, operated in low magnification ($510\times$) imaging mode, transmitted electrons continue to propagate through a set of adjustable magnetic lenses, which project the electron wave function onto the detector plane, yielding a position-resolved measurement. To reduce the background counts from electrons which have not lost energy corresponding to the emission energy of a single photon, the microscope is operated in energy-filtered TEM mode. A Gatan GIF 2001 energy filtering spectrometer is used to filter electrons that have lost 2-3 eV, corresponding to visible photon wavelengths of $\sim$600~nm - 400~nm. The majority of electrons that did not emit a photon and therefore didn't lose any energy are blocked by the filter, while electrons that did emit a visible photon pass through the filter. This filtering step significantly improves the signal-to-noise ratio and allows for the use of higher beam currents on the electron camera, which in turn reduces the overall acquisition time.

Finally, the electrons are detected and time-stamped using a Timepix3 (TP3)-based direct electron detector (Advascope ePix) with both temporal and spatial resolution and a high detection efficiency~\cite{poikela_timepix3_2014}, which has been retrofitted to the TV port of the electron spectrometer. The electron camera at the TV port is in a conjugate image plane with respect to the TEM's sample plane and the transmission mask in the photon arm. This configuration reminds the two-photon Gaussian thin-lens equation used to describe conventional ghost imaging~\cite{Pittman_Klyshko_Thin_lens_ghost}. Overall, a direct spatial mapping is obtained between the mask inserted in the photon arm and the electron camera.

The time-of-arrival differences $\tau = t_e - t_{\gamma}$ between electron arrival time, $t_e$, and CL photon arrival time, $t_{\gamma}$, are counted and visualized in a histogram, see Fig.~\ref{fig:setup}(C). To obtain the normalized correlation signal, we define the temporal cross-correlation function as $g^{(2)}(\tau) = \frac{C(\tau)}{C_{bg}}$, where $C(\tau)$ are the measured coincidence counts at delay $\tau$, and $C_{bg}$ is the average background coincidence level. In practice, $C_{bg}$ is evaluated by averaging the coincidence counts at larger delays, typically in the interval $\tau = -200\;\mathrm{ns}$, where no physical correlations are expected and only accidental coincidences contribute. This procedure ensures that $g^{(2)}(\tau) = 1$ corresponds to an uncorrelated background, while deviations $g^{(2)}(\tau) > 1$ directly reveal the electron–photon coincidence signal (see Supplemental Material for more detail).

For various reasons (e.g., TP3 detector jitter and synchronization between electron and photon detection), the temporal resolution is limited to about 50~ns. 
A pair of detection events is classified as a coincidence event if the time difference is within $\pm 25\;\mathrm{ns}$,  after accounting for a fixed offset caused by signal propagation delays, see~\cite{preimesberger_exploring_2025,preimesberger_experimental_2025} for more details.

\begin{figure*}[!t]
	\centering
\includegraphics[width=\textwidth]{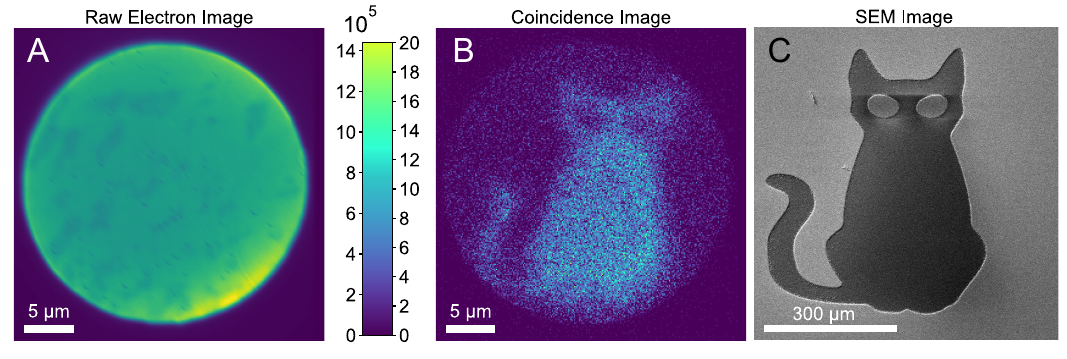} 
	\caption{\textbf{Cat ghost images}. (\textbf{A}) Raw electron beam image recorded on the TP3 electron detector. (\textbf{B}) Ghost image of the cat-shaped mask, reconstructed from electron-photon coincidences. Due to the optical system the cat-shaped mask is demagnified by a factor of $\sim19$ on the electron sample plane. (\textbf{C}) Scanning Electron Microscope image of the fabricated cat-shaped transmission mask.}
	\label{fig:cat_sem}
\end{figure*}

\section{Results}
With this experimental setup, we are able to acquire a ghost image of a cat-shaped mask (Fig.~\ref{fig:cat_sem}). The raw energy-filtered electron counts, Fig.~\ref{fig:cat_sem}(A), reveal the largely featureless silicon membrane sample, with a round intensity distribution corresponding to the size (diameter $\sim 31$ \textmu m) and shape of the incident electron beam. Inhomogeneities visible within the disc are the result of aberrations in the energy filter or contamination deposited on the membrane sample during alignment of the system. Filtering these electrons based on coincidence matching with the collected photons reveals a ghost image, Fig.~\ref{fig:cat_sem}(B), containing $N>10^5$ electron-photon coincidence events of the cat mask, Fig.~\ref{fig:cat_sem}(C). The data acquisition experiments were performed during an uninterrupted 6 hour session. The acquisition speed is primarily limited by the very low beam current (23~pA) required to prevent saturation of the time-resolved TP3 detector. With a cold-FEG source, providing higher brightness and improved monochromaticity, together with the next-generation TP4 detector~\cite{llopart2022timepix4_electron_detection}, it will be possible to operate at beam currents of $>$~1~nA, reducing the acquisition time by roughly two orders of magnitude. As already briefly discussed in the manuscript, the photon rate is also limited by an aperture that restricts the usable area of the parabolic mirror, whose surface quality and intrinsic geometry introduce spherical aberrations in imaging. Improving these optical elements will further shorten the required integration times. The post-processing procedure follows the same approach as in our previous work~\cite{preimesberger_experimental_2025}.

The mask has dimensions of $\sim 500{\times}600$~\textmu m$^2$. The eyes of the cat, which constitute the smallest resolved features, measure $\sim 70 {\times}50$~\textmu m$^2$. Fabrication of the mask is described in the Supplemental Material.
In our ghost image, we clearly resolve the features of the cat, including the eyes, ears, and tail. With a demagnification of $\sim 19 \times$ from the photon image plane to the electron sample plane, these features are on the order of a few \textmu m in the coincidence image.

In order to quantify the position resolution of the ghost imaging setup, we replace the cat mask in the photon path by a calibration target composed of absorptive grating lines of 60~\textmu m periodicity (Thorlabs R1L3S6P). With this target, we quantify the resolution of the ghost image by convolving the ideal target image with a Gaussian point spread function and fitting the model to our experimental data.
We present the resolution at the TEM sample plane, which is the plane imaged by the projective electron optics, and the plane at which the electron-photon pair originates. Using this plane allows us to directly relate the fitting results to the correlations in the underlying joint electron-photon state.

To derive the ideal target image, the straight grating lines must be back-projected through the optical system, noting that in this experiment the optical system demagnifies the lines by $16\times$ onto the TEM sample plane. The parabolic mirror introduces noticeable distortions to the straight lines, which we account for in the analytical model. We model this back-projection as described in \cite{preimesberger_experimental_2025}, leaving several parameters open for our least-absolute error fitting algorithm to optimize, such as the precise location of the sample relative to the mirror and the standard deviation of the Gaussian function. 

\begin{figure}[hb]
	\centering
    \includegraphics[width=0.5\textwidth]{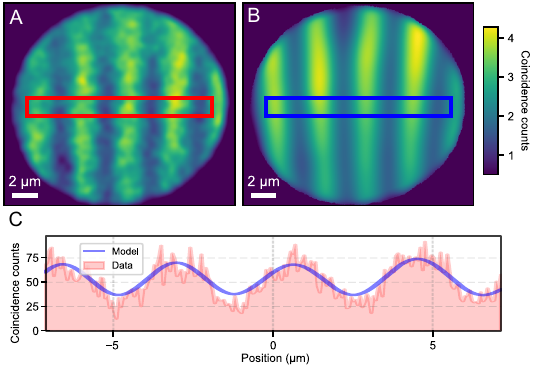}
	\caption{\textbf{Ghost image of resolution test sample}. (\textbf{A}) Ghost image of 60 \textmu m periodicity grating lines placed in the photon image plane, recorded by the TP3 electron camera, exhibiting a spatial resolution of $2.03 \pm 0.06$~\textmu m. Image was rotated and smoothed with a Gaussian filter for presentation purposes, after the fitting procedure was complete.
    (\textbf{B}) Analytical model of the grating lines back-projected through the optical system onto the TEM sample plane. (\textbf{C}) Line profiles, integrated over the transverse direction, shown for the coincidence-filtered data (red, filled) and the analytical model (blue).}
	\label{fig:data}
\end{figure}

The experimental measurements are shown in Fig.~\ref{fig:data}(A), reproducing the structure of the grating target in the electron coincidence-filtered image.
Our fitted model of the grating lines can be seen in (B); the fit exhibits a resolution of $2.03 \pm 0.06$~\textmu m full-width at half-maximum (corresponding to a standard deviation of $\Delta x =0.87 \pm 0.03$~\textmu m). 

\section{Discussion}
The spatial resolution of the quantum ghost imaging technique, in general, is restricted both by the imaging quality of the optics, and by the strength of the correlations between the pair of particles used~\cite{moreau_resolution_2018, moreau2019ghost_imaging}.

The use of electron–photon pairs instead of photon–photon pairs overcomes key limitations of the imaging capabilities of the camera arm in photonic quantum ghost imaging~\cite{moreau_resolution_2018}. The sub-atomic resolution of TEMs mitigates the wavelength-limited resolution of the photon camera arm. Additionally, direct electron detectors with near-unity detection efficiency and sub-nanosecond temporal resolution \cite{llopart2022timepix4_electron_detection} enable substantially improved image quality. Consequently, the achievable spatial resolution of the electron-photon system is primarily limited by the performance of the remaining photon arm. Although the roles of the arms could be reversed in principle, this alternate configuration is unlikely to improve the overall resolution of electron-photon ghost imaging, which continues to be limited by the photon imaging capabilities.

Additional limitations on photonic ghost imaging resolution come from the nature of photon pair production inside the commonly used spontaneous parametric down-conversion (SPDC) crystal. The uncertainty in photon production location throughout the length of the crystal and pump beam diameter leads to limitations on the phase-matching and spatial resolution of the system~\cite{asban_quantum_2019}. Reducing the pump beam diameter may improve phase-matching conditions in far-field ghost imaging setups, but has been shown to increase the separability of the photon-photon state~\cite{saleh_duality_2000}, reducing the overall spatial resolution. 
Attempts to optimize one parameter often degrade another, making high-resolution ghost imaging challenging~\cite{balakin_quantum_2022}.

Transition radiation, on the other hand, is an interface phenomenon which occurs when a charged particle enters or exits a dielectric medium~\cite{frank_transition_1966}. In the TEM, this emission is localized to the position where the electron enters or exits the silicon membrane. This removes the concerns regarding the thickness of the pair production zone.

Additionally, energy and transverse momentum conservation during the emission process directly link the electron scattering angles to the transverse momentum distribution of the emitted photons~\cite{preimesberger_exploring_2025}, enforcing a strong transverse momentum anti-correlation $\vec{p}_{\gamma\perp}=\vec{p}_{\text{e}\perp}$. This results in a mirror-like transverse momentum correlation, analogous to conventional SPDC-based ghost imaging~\cite{Rubin_Mirror_like_behaviour, KLYSHKO1988299}. Despite this similarity, electrons and photons are fundamentally different particles with vastly different masses and dispersion relations. For 200 keV electrons, transition radiation in the 2 – 3 eV range is emitted at typical photon angles of $\sim35^\circ -50^\circ$, while the corresponding electron scattering angles are $\sim2-4$~\textmu rad~\cite{frank_transition_1966}. Consequently, the scattering geometry is highly asymmetric, resembling the use of non-degenerate particles in SPDC-based ghost imaging (e.g. described in~\cite{Rubin_Non-degenerate_ghost}).

\section{Conclusion}
In this work, we have demonstrated the first realization of quantum ghost imaging of masks imprinted with two-dimensional patterns using strongly correlated electron–CL photon pairs. By developing a custom free-space CL collection system and implementing time-resolved coincidence detection, we achieved complex image reconstruction with $\Delta x =0.87 \pm 0.03$~\textmu m on the electron-photon pair source plane.
Compared to our previous work ($\Delta x \approx 1.45$~\textmu m)~\cite{preimesberger_experimental_2025}, we show here an increase in the spatial resolution of almost 50\%, which we attribute largely to improvements made to the photon optical system, reducing the contribution of aberrations and surface imperfections of the parabolic mirror.
With this upgrade, we approach the ideal conditions required for future demonstrations of quantum entanglement, such as, EPR-states ~\cite{einstein1935einsteinpodolskyrosen, reidColloquiumEinsteinPodolskyRosenParadox2009} and Bell states~\cite{bell1964einstein}, and for exploiting these states in discrete-variable protocols, as proposed in~\cite{Rembold2025_TEM_entanglement}.

Our approach also opens exciting opportunities for programmable and dynamic ghost imaging and beam-shaping based on post-selection. Adaptive masks could be generated using a digital mirror device (DMD)~\cite{mirhosseiniRapidGenerationLight2013} or implemented through spatially resolved photon detection. Such schemes would enable the post-selection shaping of electron wave functions, optimizing for quantum-enhanced sensing. This level of control could allow researchers to probe previously inaccessible features with unprecedented precision, harnessing engineered electron wave functions for maximal sensitivity to parameters of interest in the sample~\cite{yu_recent_2015, yu_wavefront_2022, haslinger2024spin}. 
Electron-photon ghost imaging holds the possibility to overcome several disadvantages of current amplitude-shaping techniques, which largely rely on diffractive gratings that produce limited beam separation on the sample plane~\cite{agarwal_nanofabricated_2017, yasin_path-separated_2018}. 
Moreover, implementing electron-photon-based wavefront shaping before illuminating the sample allows for heralding setups~\cite{signorini_-chip_2020}, for contrast improvement and low-dose imaging~\cite{johnson_single-pixel_2022}, or for improved state preparation for interferometric measurements~\cite{yasin_path-separated_2018, turner_interaction-free_2021, henke_ropers2025entanglement}. 

The correlation-based nature of ghost imaging inherently provides enhanced robustness against noise~\cite{li2021enhancing}, making it particularly valuable for low-signal or interaction-constrained environments, where conventional imaging techniques are limited~\cite{saldin2016ghost}. Future extensions of our setup could merge electron-photon pair ghost imaging with advanced photonic techniques, including phase-object imaging~\cite{abouraddy_entangled-photon_2004}, optical coherence tomography~\cite{teichVariationsThemeQuantum2012}, ghost holography~\cite{songExperimentalObservationOnedimensional2013}, or macroscale 3D ghost LiDAR~\cite{gong_three-dimensional_2016, wang_airborne_2018}.

\section*{Funding Information and Acknowledgments}
The authors thank Leonie Walrath, Larissa Putz and the USTEM for fruitful discussions. PH, AP, SB, IB thank the Austrian Science Fund (FWF): Y1121, P36041, P35953.  IB also acknowledges support during manuscript revision from the Austrian Science Fund (FWF) under grant 10.55776/ESP7270024. This project was supported by the ESQ-Discovery Program 2020 "A source for correlated electron-photon pairs" and the FFG-project AQUTEM. 

\bibliography{bib}
\clearpage
\onecolumngrid

\setcounter{figure}{0}
\renewcommand{\thefigure}{S\arabic{figure}}

\section{Supplemental Material}
\subsection{Temporal cross-correlation function}
In quantum optics, correlation functions measure the correlations in the electric field $E(\vec{r}_1,t), E(\vec{r}_2,t+\tau)$ as a function of the time difference $\tau$. While the first order coherence function is defined using the electric fields, the second order correlation function $g^{(2)}(\tau)$ can be expressed in terms of the field intensities: 
\begin{equation*}
    g^{(2)}(\tau )=\frac{\left\langle I_1(t)I_2(t+\tau )\right\rangle }{\langle I_1(t)\rangle\langle I_2(t)\rangle}
\end{equation*}
with $I_1$ at position $\vec{r}_1$ and $I_2$ at position ($\vec{r}_2$). It can thus be applied straightforwardly to the case of electron-photon pairs. We choose this nomenclature in order to highlight the parallels to the familiar two-photon case.

\subsection{Parabolic mirror manufacture}
Our CL photon extraction setup relies on a custom-built miniaturized parabolic mirror designed to fit within the spatial constraints of the TEM polepiece, while maintaining electron beam access to the sample region. 
The mirror design parameters were modeled and optimized such that the focal point of the mirror could coincide with typical imaging conditions in the TEM.
The parabolic mirror was fabricated from aluminum using a computer-numerical-control (CNC) milling machine, as described below, and can be mounted on a custom single tilt TEM sample holder tip. 

\begin{figure}[h]
    \centering
    \includegraphics[width=0.45\linewidth]{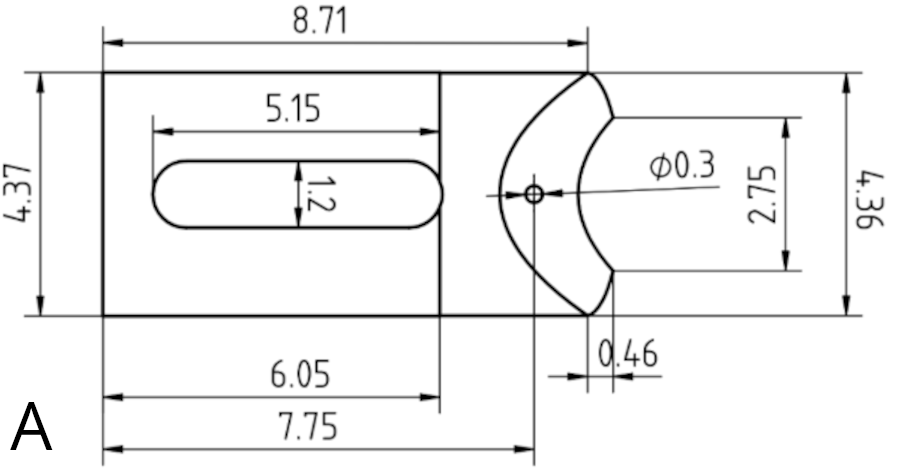}
    \includegraphics[width=0.45\linewidth]{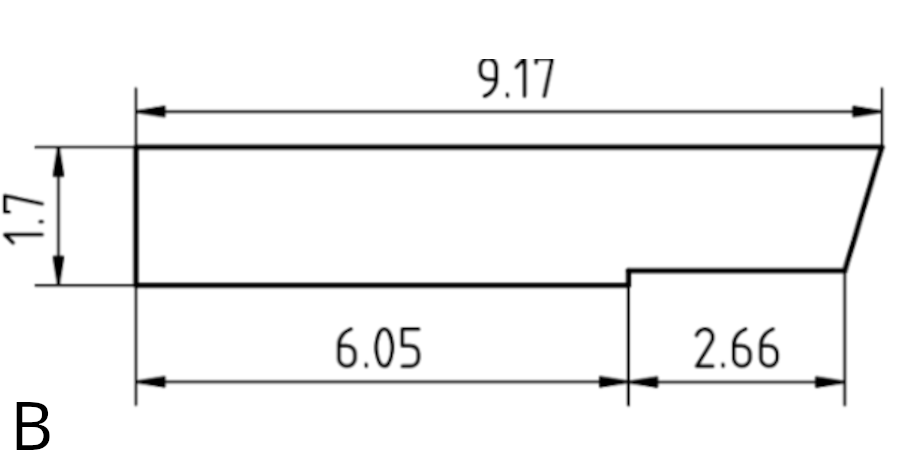}
    \caption{Technical drawing from bottom (\textbf{A}) and side perspectives (\textbf{B}) of the parabolic mirror, showing the outline dimensions. The curvature was chosen to match a focal length of $f=\SI{750}{\micro\meter}$ with an $\mathrm{NA} \approx 0.58$. All dimensions specified in millimeters.}
    \label{fig:mirror_dimensions}
\end{figure}

\begin{figure}[bp]
    \centering
    \includegraphics[width=0.45\linewidth]{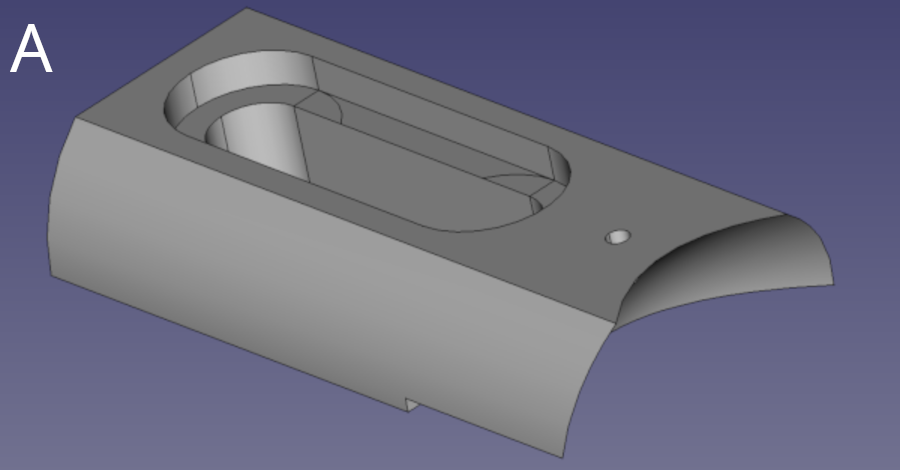}
    \includegraphics[width=0.45\linewidth]{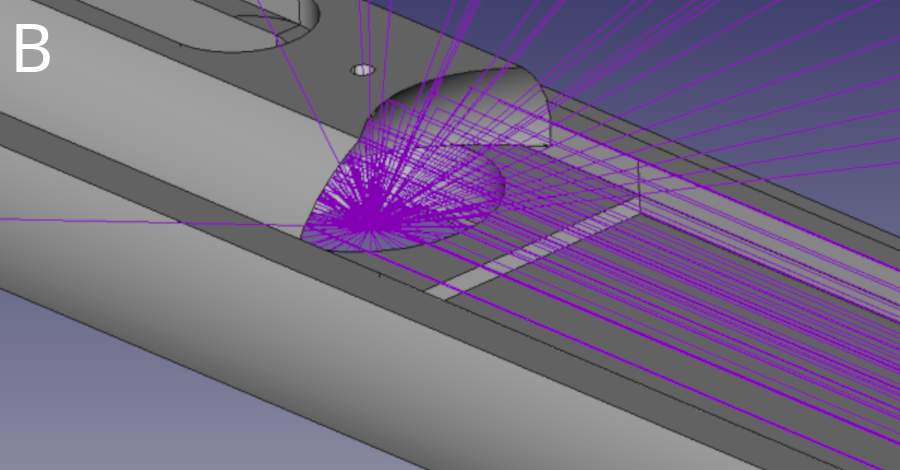}
    \caption{(\textbf{A}) Rendering of the parabolic mirror design in FreeCAD. (\textbf{B}) Purple rays trace the path of photons emitted from a sample mounted at the focal point of the mirror on the TEM holder tip. The simulation was performed using the optics workbench of FreeCAD.}
    \label{fig:mirror_design_overview}
\end{figure}

The mirror (numerical aperture: $\mathrm{NA} \approx 0.58$) was designed to couple light from the IR to the UV range out of the TEM by directing the light through an optical window installed near the polepiece region of the electron column. The mirror is attached to the upper side of our customized TEM sample holder, with a focal point ($f = \SI{750}{\micro\meter}$) set to coincide with a typical sample mounting height in the single tilt holder.
Since the mirror also covers the sample along the electron beam path, we incorporate a \SI{300}{\micro\meter} diameter beam hole into the contoured mirror. The mirror spans a height of \SI{1.7}{\mm} and a width of \SI{4.37}{\mm} (see Fig.~\ref{fig:mirror_dimensions}), in order to optimize the optical coverage inside the limited space within the TEM's objective polepiece. The mirror design was created using FreeCAD design software; a 3D rendering of the design is presented in Fig.~\ref{fig:mirror_design_overview}(A).

We further validated our design using FreeCAD's optics workbench to simulate photon ray paths. This environment allowed us to perform raytracing at different sample positions and different sample holder alignments (see Fig.~\ref{fig:mirror_design_overview}(B)), as well as enabling a coarse simulation of external lens alignment.

\begin{figure}[htbp]
    \centering
    \includegraphics[width=0.45\linewidth]{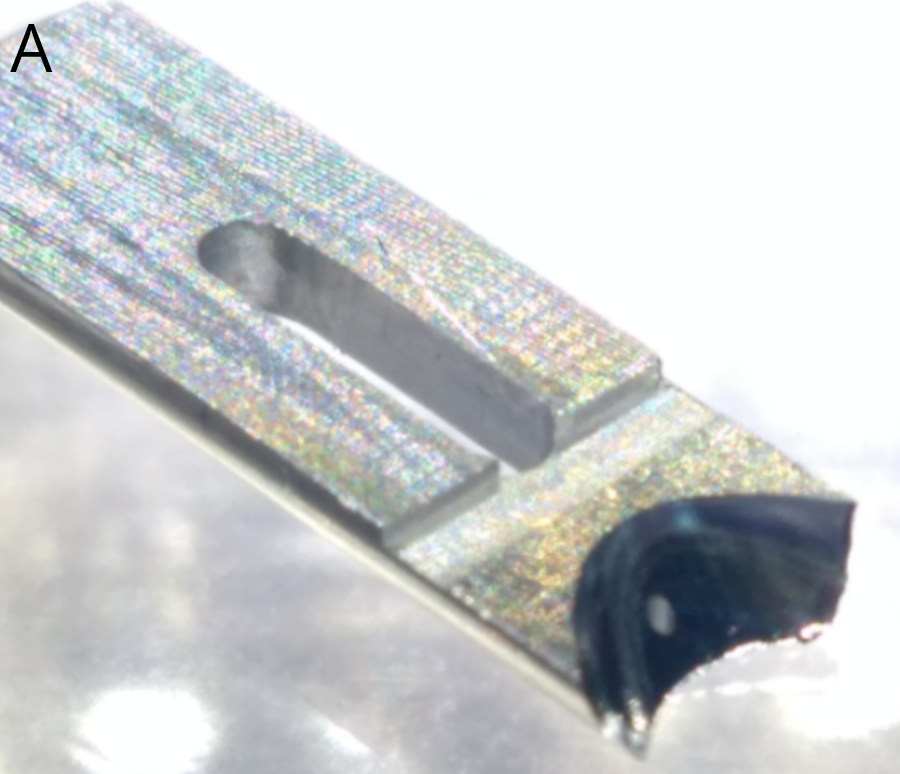}
    \includegraphics[width=0.45\linewidth]{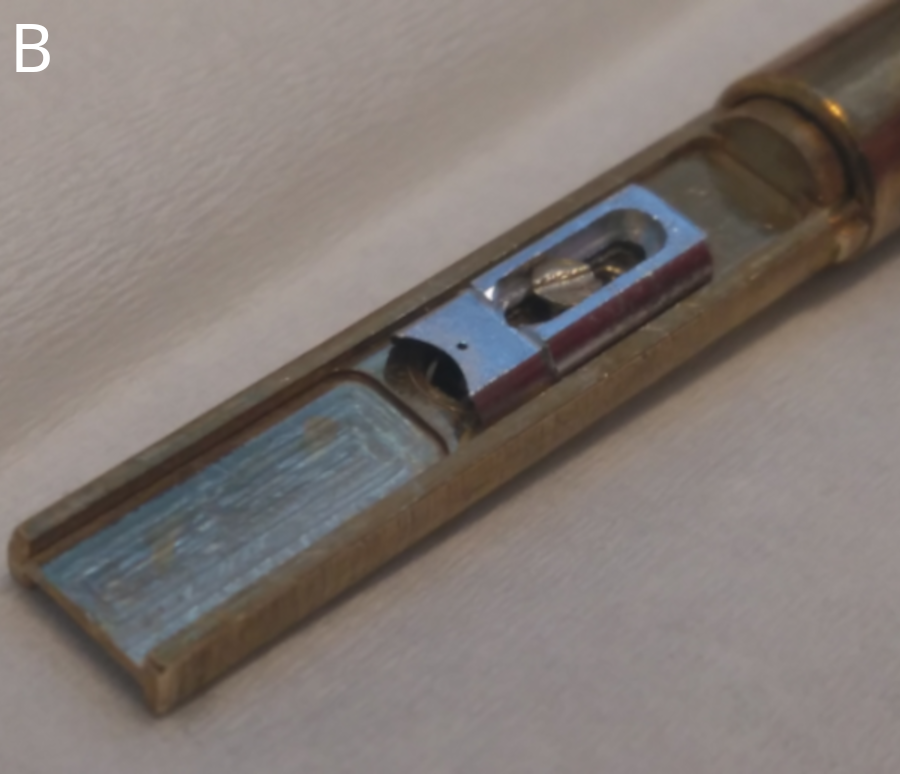}

    \includegraphics[width=0.45\linewidth]{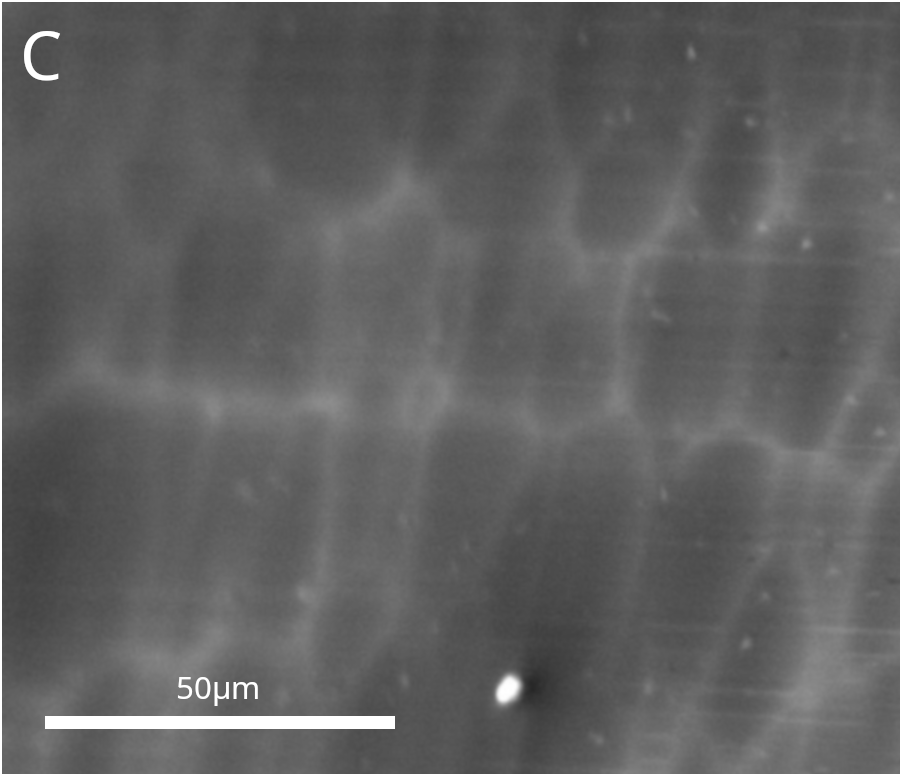}
    \includegraphics[width=0.45\linewidth]{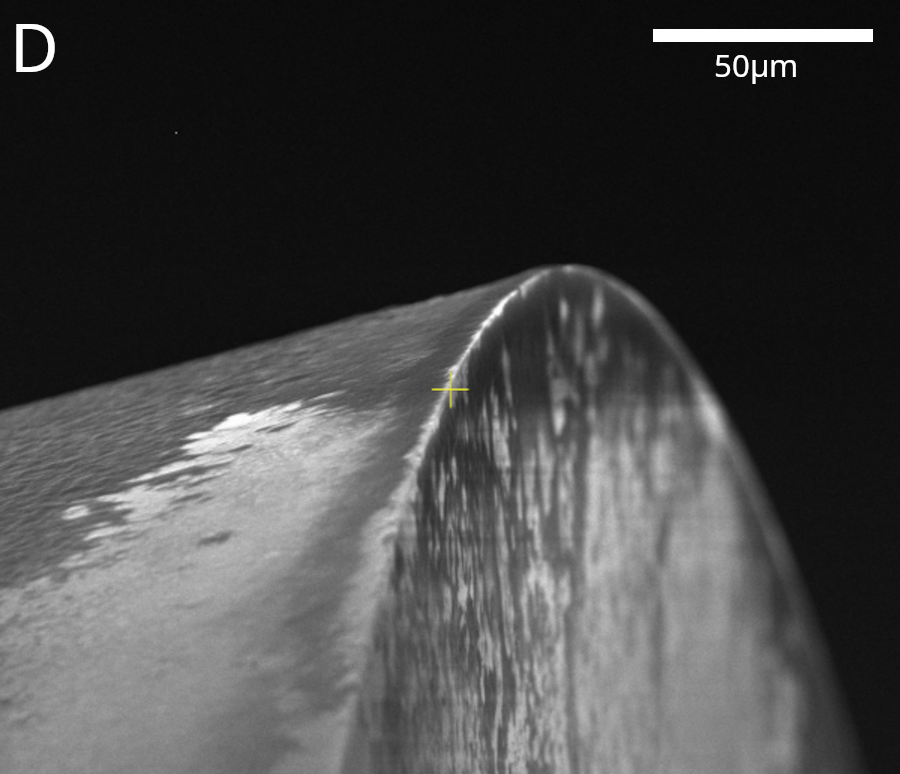}
    \caption{(\textbf{A}) Photograph of the mirror, upside-down to reveal the reflective surface. (\textbf{B}) The mirror mounted on the custom sample holder tip. (\textbf{C}) SEM image of the mirror surface, showing the surface quality. The thin horizontal lines originate from the line-by-line machining of the surface. The larger scale pattern visible on the surface originates from the grain size of the 6060 aluminum alloy\protect\cite{KAYSER20106568}. Except for micrometer sized sporadic dirt particles on the surface (e.g. the white spot near the bottom), remaining features such as the horizontal lines and scratches left by the mill have sizes on the order of \SI{130}{\nano\meter} and below. (\textbf{D}) SEM image of the cutting edge of the monocrystalline diamond ball end mill that was used for final surface milling. We utilized the SEM measurements to determine the exact curvature of the tools to generate the toolpaths for the CNC mill.}
    \label{fig:mirror_mounted_sem}
\end{figure}

We chose to use aluminum for the mirror due to its very broadband optical response~\cite{adachi2012handbook}, easy machinability, and compatibility with diamond tools.
We tested aluminum alloys with a range of hardness values, namely 6060, 5083, 6082, and 7075 aluminum alloy. Comparative testing revealed that the softer 6060 alloy performed best during manufacturing. This softer alloy experienced more smearing during the machining process, which aided in reducing surface errors compared to the other, more brittle alloys.
Therefore, we chose 6060 Aluminum alloy for the mirror material.
In the visible range that we used for initial characterization of the mirror we expect a reflectivity of about 85\% in the optimal case. We verified the reflectivity to be around 85\% for \SI{650}{\nano\meter}  light utilizing a Thorlabs PM100D power meter. See Fig.~\ref{fig:mirror_mounted_sem}(A) for an image of the mirror.

In order to attach the mirror to the TEM holder, the design of the mirror body includes a slot, which is used with a metric M\,1.6\,$\times$\,0.35 steel screw to fasten the mirror to a custom-designed brass sample holder tip (Fig.~\ref{fig:mirror_mounted_sem}(B)). The slot design allows transverse movement of the mirror with respect to the sample. The torque transferred from the screw to the mirror introduces stress into the mirror; the expected deformation has been deemed insignificant for the setup.

Since the required curvature of the parabolic surface is inaccessible to traditional machining facilities, such as grinding, we performed direct milling of the surface. As our milling platform we utilize the Krause Mauser Präzoplan 300 CNC machine~\cite{jaumann2012prazoplan}. The machine utilizes an area-guided $XY$ axis, to prevent stacking errors (i.e. the propagation of errors from one axis onto the other axis), with aerostatic guides. The aerostatic guides utilize a constant pressure air film between the stationary and moving part of the guides to reduce friction and wear as well as preventing stick-slip effects. The $XY$ table is integrated into a $Z$ axis portal. In its original configuration the machine achieves positioning repeatability down to \SI{30}{\nano\meter}, limited by the effects of friction on the aerostatic guides~\cite{jaumann2012prazoplan}, and an absolute positional accuracy of \SI{1}{\micro\meter} utilizing glass measurement sticks at the guides. When doing \SI{100}{\nano\meter} steps the machine showed less than \SI{10}{\nano\meter} deviations~\cite{jaumann2012prazoplan}. In addition to the glass measurement sticks mounted on the guides of the machine, we utilized an additional optical measurement grid for enhanced feedback on the position. This grid has previously been used to characterize the machine. It is mounted directly below the tool center point.

The CNC machine is controlled via a Heidenhain control system. The manufacturing files were exported from FreeCAD as STEP files and prepared using Siemens NX CAD/CAM software as well as custom postprocessing scripts. We utilized the dimensions gathered via scanning electron microscope (SEM) measurements to adjust the toolpath to the end mill curvature (see Fig.~\ref{fig:mirror_mounted_sem}(D)).

The machining process employed a comprehensive set of precision cutting tools specifically optimized for aluminum processing. Initial roughing operations were performed using solid carbide (VHM) flat end mills with diameters of \SI{4}{\mm} and \SI{2}{\mm}, respectively, manufactured by Holex (Hoffman Group). A standard \SI{0.3}{\mm} solid carbide drill was utilized for creating the beam hole with high dimensional accuracy. Following this, semi-finishing passes were executed using a \SI{2}{\mm} full-radius solid carbide end mill, also sourced from Holex. The solid carbide tools were periodically cleaned by soaking them in caustic soda to prevent clogging of the flutes. The final precision finishing passes were accomplished with a ball shaped \SI{1}{\mm} full-radius monocrystalline diamond (MKD) end mill, manufactured by Paul Horn, ensuring superior surface quality and dimensional tolerances (see Fig.~\ref{fig:mirror_mounted_sem}(C),(D)).

As a postprocessing step the mirror was cleaned in an ultrasonic bath at \SI{60}{\degreeCelsius} in isopropyl alcohol to remove residual oil from the mist minimum quantity lubrication process. Finally, the mirror was polished manually with increasingly fine abrasive pastes, down to a \SI{20}{\nano\meter} diamond suspension, before cleaning again and use in the experiment.

\subsection{Cat-shaped mask}
The cat-shaped mask was fabricated using a focused ion beam scanning electron microscope (FIB-SEM). A glass substrate (microscope slide) was coated with a $\sim 2.5$ \textmu m thick polycrystalline Ag layer by sputter deposition. The mask design was created in Adobe Photoshop (version: 26.10.0) and exported as a bitmap file ($4000{\times}4000$ pixels) compatible with the FIB-SEM control software. The mask guided the selective milling of the Ag layer with the ion beam, resulting in the cat-shaped mask shown in the SEM image in Fig.~\ref{fig:cat_sem}(C). Charging effects are visible in the SEM image due to the removal of the conductive Ag layer, which exposed the underlying insulating glass substrate. 

\end{document}